\documentclass[conference]{IEEEtran}
\ifCLASSINFOpdf
\else
\fi

\usepackage{mathrsfs}
\usepackage{caption2}
\usepackage{url}
\usepackage{stfloats}
\usepackage{fixltx2e}
\usepackage{eqparbox}
\usepackage{mdwmath}
\usepackage{mdwtab}
\usepackage{array}
\usepackage{algorithmic}
\usepackage[cmex10]{amsmath}

\usepackage{graphicx}
\begin{document}
\title{Outage Performance of AF-based Time Division Broadcasting Protocol in the Presence of Co-channel Interference}

\author{\IEEEauthorblockN{Xiaochen Xia, Youyun Xu, Kui Xu, Dongmei Zhang, and Ning Li}
\IEEEauthorblockA{Institute of Communication Engineering, PLA University of Science and Technology\\
Nanjing 210007, P. R. China\\
Email: Xia1382084@gmail.com }}


%Telephone: (800) 555--1212, Fax: (888) 555--1212}


\maketitle
\

\begin{abstract}
In this paper, we investigate the outage performance of time division broadcasting (TDBC) protocol in independent but non-identical Rayleigh flat-fading channels, where all nodes are interfered by a finite number of co-channel interferers. We assume that the relay operates in the amplified-and-forward mode. A tight lower bound as well as the asymptotic expression of the outage probability is obtained in closed-form. Through both theoretic analyses and simulation results, we show that the achievable diversity of TDBC protocol is zero in the interference-limited scenario. Moreover, we study the impacts of interference power, number of interferers and relay placement on the outage probability. Finally, the correctness of our analytic results is validated via computer simulations.
\end{abstract}


% \fi
\IEEEpeerreviewmaketitle

\section{Introduction}
Recently, two-way relaying (TWR) or bi-directional relaying has emerged as a powerful technique to improve the spectral efficiency of wireless network \cite{IEEEhowto:Rankov}. A number of relaying protocols have been proposed, such as amplify-and-forward (AF), decode-and-forward (DF) and compress-and-forward (CF). For AF relaying, two popular TWR protocols are analog network coding (ANC) \cite{IEEEhowto:Katti_1}, which requires two time slots to complete the information exchange between two terminal nodes, and TDBC \cite{IEEEhowto:Kim_1}, which needs three time slots. However, TDBC protocol can use the direct link between two terminals even under a half-duplex constraint \cite{IEEEhowto:Kim_1}\cite{IEEEhowto:Zaeri-Amirani}, thus can provide higher diversity gain.

Several previous works have investigated the TWR network using TDBC for Rayleigh fading channels, in which relay and terminals are only perturbed by additive white Gaussian noise (AWGN) \cite{IEEEhowto:Yi}-\cite{IEEEhowto:Ju}. The outage performance of AF-based TDBC protocol in Rayleigh fading channels was analyzed in \cite{IEEEhowto:Yi}\cite{IEEEhowto:Ding} and the diversity-multiplexing tradeoff (DMT) was also obtained. In \cite{IEEEhowto:Ju}, the authors considered relay selection scheme for TDBC protocol and analyzed the outage performance with optimal relay selection. However, signals of terminals (or relay) are often corrupted by co-channel interference (CCI) from other sources that share the same frequency resources in wireless networks \cite{IEEEhowto:Hoeher}. Moreover, for the wireless scenarios with dense frequency reuse,
co-channel interference may dominate the AWGN. Therefore, it is necessary to take the effect of CCI into serious consideration in the analysis and design of the practical TDBC protocol. In \cite{IEEEhowto:Ikki}, the performance of ANC protocol corrupted by equal-power interferers was studied, where closed-form expressions of the average bit error rate and outage probability were presented. Outage probability of the cooperative relaying using DF protocol with CCI has been analyzed in \cite{IEEEhowto:Lee}. However, for AF-based TDBC protocol, the effect of CCI is still unknown.
\begin{figure}[t]
\centering
\includegraphics[width=6.2cm]{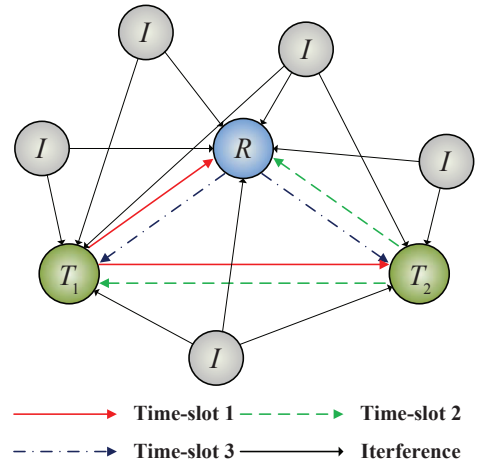}
\caption{The TDBC with a finite number of co-channel interferers, where $I$ denotes the co-channel interferer.} \label{fig:graph}
\end{figure}

In this work, we study the AF-based TDBC protocol where all the nodes (terminals and relay) are interfered by a finite number of co-channel interferers in independent but non-identical Rayleigh flat-fading channels. The system model is described in the next section. In section III, the CDF of the upper bounded SINR is analyzed. Based on the results, a lower bound as well as asymptotic expression of outage probability is obtained. In section IV, the effects of interference power, number of interferers and relay placement on the outage probability are studied.

\section{System model}
We study the TWR network which consists of two terminals and a relay node, as shown in Fig. 1, in which terminal $T_1$ and terminal $T_2$ exchange statistically independent messages with the help of a relay $R$. Each node is equipped with a single antenna and operates in the half-duplex mode, that is, a node cannot transmit and receive simultaneously.

The TDBC protocol can be achieved within three time slots, that is, terminal $T_1$ transmits during the first time slot, while $T_2$ and $R$ listen. In time slot 2, $T_2$ transmits while $T_1$ and $R$ listen. It is assumed that both terminals and the relay are interfered by a finite number of co-channel interferers. Denoting $L_R$, $L_1$ and $L_2$ as the total numbers of interferers that affect node $R$, $T_1$ and $T_2$, respectively, the received signals at the relay and $T_i$ during the first two time slots are expressed as\\
\begin{equation}
\begin{aligned}
{y_{iR}} = \sqrt {{E_i}} {h_i}{S_i} + \sqrt {{E_I}} \sum\limits_{k = 1}^{{L_R}} {{d_{R,k}}I_{R,k}^i}  + {n_{iR}},\\
{y_{ji}} = \sqrt {{E_j}} {h_0}{S_j} + \sqrt {{E_I}} \sum\limits_{k = 1}^{{L_i}} {{d_{T_i,k}}I_{{T_i},k}^j}  + {n_{ji}},\\
\end{aligned}
\end{equation}
respectively, where $i,j \in \left\{1,2 \right\}$ and $i\ne j$. $E_i$ and $E_I$ denote the transmit powers of $T_i$ and interferers, respectively. $h_1$, $h_2$ and $h_0$ represent the channel coefficients belonging to the links $T_1\to R$, $T_2\to R$ and $T_1\to T_2$, respectively. The channel reciprocity is assumed. Moreover, $d_{N,k}$ indicates the channel coefficient of link between node $N$ and the $k$th interferer that affects $N$, where $N\in\{T_1,T_2,R\}$. All links are assumed to be independent but non-identical Rayleigh flat-fading. $S_i$ denotes the unit-power symbol transmitted by $T_i$. $I_{N,k}^{m}$ indicates the unit-power interference signal of $k$th interferer that affects node $N$ during the $m$th time slot, where $N\in\{T_1,T_2,R\}$ and $m\in\{1,2,3\}$. Finally, $n_{iR}/n_{ji}$ denote the AWGN and $n_{iR}/n_{ji}\sim{\cal C}{\cal N}\left( {0,1}\right)$.

In time slot 3, $R$ transmits the combined information to terminals $T_1$ and $T_2$. The combined signal to be transmitted by $R$ can be written as $S_R = {\cal A}_1 y_{1R} + {\cal A}_2 y_{2R}$. ${\cal A}_1$ and ${\cal A}_2$ denote combining coefficients which can be determined as {follow\footnotemark[1]} \cite{IEEEhowto:Louie}:
{\footnotetext[1]{As in \cite{IEEEhowto:Ikki}, here it is assumed that $R$ knows the channel gains of links $T_1\to R$ and $T_2\to R$, and the total interference power (instantaneous) at $R$. Moreover, it is assumed that $T_i$ knows the channel gains of links $T_1\to R$, $T_2\to R$, $T_1\to T_2$ as well as the total interference powers (instantaneous) at $R$ and $T_i$.}}
\begin{equation}
{{\cal A}_i} \!=\! \sqrt {\frac{{{\omega _i}}}{{{\omega _1}E_1{{\left| {{h_1}} \right|}^2} + {\omega _2}E_2{{\left| {{h_2}} \right|}^2} + {E_I}\sum_{k = 1}^{{L_R}} {{{\left| {{d_{R,k}}} \right|}^2} \!+\! 1} }}},
\end{equation}
where $i\in \{1, 2\}$. $\omega _i\in \left[0,1\right]$ is the power allocation number and ${\omega _1} + {\omega _2} = 1$. Then the received signal at terminal $T_i$ during the third time slot can be written as
\begin{equation}
\begin{array}{l}
{y_{Ri}} = \sqrt {E_r} {h_i}{S_R} + \sqrt {{E_I}} \sum\limits_{k = 1}^{{L_i}} {{d_{T_i,k}}} {I_{T_i,k}^{3}} + {n_{Ri}} \\
 = \!\!\sqrt{E_rE_1}{{\cal A}_1}{h_i}{h_1}{S_1}\!+ \!\sqrt{E_rE_2}{{\cal A}_2}{h_i}{h_2}{S_2}\! +\! \sqrt {{E_I}}\! \sum\limits_{k = 1}^{{L_i}} \!{{d_{T_i,k}}} {I_{T_i,k}^{3}} \\
 + \sqrt {E_r{E_I}} {h_i}{{\cal A}_1}\sum\limits_{k = 1}^{{L_R}} {{d_{R,k}}{I_{R,k}^1}}  + \sqrt {E_r{E_I}}{h_i} {{\cal A}_2}\sum\limits_{k = 1}^{{L_R}} {{d_{R,k}}{I_{R,k}^2}}   \\
 + \sqrt {E_r} {{\cal A}_1}{h_i}{n_{1R}} + \sqrt {E_r} {{\cal A}_2}{h_i}{n_{2R}} + {n_{Ri}},\\
\end{array}
\end{equation}
where ${n_{Ri}}\sim{\cal C}{\cal N}\left( {0,1}\right)$ is the AWGN and $E_r$ is the transmit power of $R$. In the following, we assume equal power {allocation\footnotemark[2]} between $T_1$, $T_2$ and ${R}$, i.e., $E_1=E_2=E_r=E$. Since $T_i$ knows its own transmitted symbols, it can cancel the self-interference component in ${y_{Ri}}$. Therefore, after performing maximal-ratio combining, the instantaneous SINR at terminal $T_i$ can be expressed as in (4), where $i,j\in \{1,2\}$ and $i\ne j$. By substituting the expressions of ${\cal A}_1$ and ${\cal A}_2$ into (4) and performing some manipulations, ${\gamma _{{T_i}}}$ can be rewritten as \cite{IEEEhowto:Ikki}
{\footnotetext[2]{Similar as in \cite{IEEEhowto:Ding} and \cite{IEEEhowto:Lee}, the assumption of equal power allocation dose not make the analysis in this work lose generality because the variances of the channel coefficients between $T_1$, $T_2$ and $R$ may be different.}}
\newcounter{mytempeqncnt}
\begin{figure*}
\normalsize
\setcounter{mytempeqncnt}{\value{equation}}
% Set the equation number to one less than the one
% desired for the first equation here.
% The value here will have to changed if equations
% are added or removed prior to the place these
% equations are referenced in the main text.\int\limits_{}^{} {}
%\quad\;
\setcounter{equation}{3}
\begin{equation}
{\gamma _{{T_i}}} = \frac{{E{{\left| {{h_0}} \right|}^2}}}{{{E_I}\sum_{k = 1}^{{L_i}} {{{\left| {{d_{T_i,k}}} \right|}^2}}  + 1}}
+ \frac{{{\cal A}_j^2{E^2}{{\left| {{h_1}} \right|}^2}{{\left| {{h_2}} \right|}^2}}}{{\left( {{\cal A}_1^2 + {\cal A}_2^2} \right)E{{\left| {{h_i}} \right|}^2}\big( {{E_I}\sum_{k = 1}^{{L_R}} {{{\left| {{d_{R,k}}} \right|}^2} + 1} } \big) + {E_I}\sum_{k = 1}^{{L_i}} {{{\left| {{d_{T_i,k}}} \right|}^2}}  + 1}},
\end{equation}
\setcounter{equation}{\value{mytempeqncnt}}
\hrule
\end{figure*}
\setcounter{equation}{4}
\begin{equation}
{\gamma _{{T_i}}} \approx {\gamma _{{T_i},D}} + \frac{{{\gamma _{{T_i},1}}{\gamma _{{T_i},2}}}}{{{\gamma _{{T_i},1}} + {\gamma _{{T_i},2}}}},
\end{equation}
where ${\gamma _{{T_i},D}}=\frac{{E{{\left| {{h_0}} \right|}^2}}}{{{E_I}\sum_{k = 1}^{{L_i}} {{{\left| {{d_{T_i,k}}} \right|}^2}}  + 1}}$ is the received SINR of link $T_j\to T_i$. Moreover, $\gamma _{{T_i},1}$ and $\gamma _{{T_i},2}$ are given by
\begin{equation}
\begin{aligned}
& \gamma _{{T_i},1} = {\frac{{E{{\left| {{h_i}} \right|}^2}}}{{{E_I}\sum_{k = 1}^{{L_i}} {{{\left| {{d_{T_i,k}}} \right|}^2} + 1} }}}\\
& {}\gamma _{{T_i},2} = \frac{{{\omega _j}E{{\left| {{h_j}} \right|}^2}}}{{{E_I}\sum_{k = 1}^{{L_R}} {{{\left| {{d_{R,k}}} \right|}^2} + } {\omega _i}{E_I}\sum_{k = 1}^{{L_i}} {{{\left| {{d_{T_i,k}}} \right|}^2} + {\omega _i} \!+\! 1} }}.
\end{aligned}
\end{equation}
\section{Outage Probability Analysis}
\subsection{Lower Bound of the Exact Outage Probability}
In this section, the outage probability of the AF-based TDBC protocol in the presence of CCI is studied. For brevity of analysis and without loss of generality, we focus on the outage probability at terminal $T_1$ in the rest of this work. By definition, the outage event occurs when the mutual information at $T_1$ falls below the target rate $R_t$, or equivalently, the output SINR at $T_1$ is below the target SINR $\varphi$. Therefore, the outage probability at terminal $T_1$ can be written as
\begin{equation}
P_{{T_1}}^{OUT}\left( {{R_t}} \right) = {\rm{Pr}}\left( {{I_{{T_1}}} < {R_t}} \right) = {F_{{\gamma _{{T_1}}}}}\left( \varphi  \right),
\end{equation}
where $I_{{T_1}} = \frac{1}{3}\log \left( {1 + {\gamma _{{T_1}}}} \right)$ indicates the mutual {information\footnotemark[3]} at terminal $T_1$, and $\varphi=2^{3R_t}-1$. ${F_\Psi }\left( \eta  \right)$ represents the CDF of random variable (RV) $\Psi$. However, it is very difficult to obtain the exact expression of ${F_{{\gamma _{{T_1}}}}}\left( \varphi  \right)$ in closed-form. To circumvent this obstacle, we introduce a tight upper bound on the received SINR at $T_1$ by employing a widely used inequality, i.e., ${{{\nu _1}{\nu _2}} \mathord{\left/
 {\vphantom {{{\nu _1}{\nu _2}} {\left( {{\nu _1} + {\nu _2}} \right)}}} \right.
 \kern-\nulldelimiterspace} {\left( {{\nu _1} + {\nu _2}} \right)}} \le \min \left\{ {{\nu _1},{\nu _2}} \right\}$, where $\nu _1$ and $\nu _2$ are positive numbers, then we shall have
{\footnotetext [3]{Herein, three time slots required for the TDBC protocol account for the pre-log factor of 1/3.}}
\begin{equation}
{\gamma _{{T_1}}} \le \gamma _{{T_1}}^{UB} = {\gamma _{T_1,D}} + \min \left\{ \gamma _{{T_1},1},\gamma _{{T_1},2} \right\}.
\end{equation}
Next, we will determine the CDF of the upper bounded SINR. For convenience of analysis, letting $X\buildrel \Delta \over =E{\left| {{h_1}} \right|^2}$, $Y \buildrel \Delta \over = E{\left| {{h_2}} \right|^2}$, $Z \buildrel \Delta \over = E{\left| {{h_0}} \right|^2}$, $S \buildrel \Delta \over = {E_I}\sum_{k = 1}^{{L_R}} {{{\left| {{d_{R,k}}} \right|}^2}}$ and $T \buildrel \Delta \over = {E_I}\sum_{k = 1}^{{L_1}} {{{\left| {{d_{T_1,k}}} \right|}^2}}$. Note that $X$ , $Y$ and $Z$ are exponential RVs with means $E{\Omega _{1}}$, $E{\Omega _{2}}$ and $E{\Omega _{0}}$, respectively, where ${\Omega _{i}}$ indicates the variance of $h_i$, $i \in \{0,1,2\}$. Letting $\gamma _{m}= \min \left\{ \gamma _{{T_1},1},\gamma _{{T_1},2} \right\}$, then with the help of total probability theorem, the CDF of $\gamma _{{T_1}}^{UB}$ conditioned on $S$ and $T$ can be written as in (9),
\begin{figure*}
\normalsize
\setcounter{mytempeqncnt}{\value{equation}}
% Set the equation number to one less than the one
% desired for the first equation here.
% The value here will have to changed if equations
% are added or removed prior to the place these
% equations are referenced in the main text.\int\limits_{}^{} {}
%\quad\;
\setcounter{equation}{8}
\begin{equation}
\begin{aligned}
{F_{\gamma _{{T_1}}^{UB}\left| {\left\{ {S,T} \right\}} \right.}}\left( \varphi  \right) &= 1 - \Pr \left( {\left. {{\gamma _{{T_1},D}} > \varphi } \right|T} \right) - \Pr \left( {\left. {{\gamma _{{T_1},D}} < \varphi ,{\gamma _m} > \varphi  - {\gamma _{{T_1},D}}} \right|S,T} \right) \\
&{} =1 - \int_\varphi ^\infty  {{f_{{\gamma _{{T_1},D}}|T}}\left( r \right)dr}  - \int_0^\varphi  {\left( {1 - {F_{{\gamma _{m\left| {\left\{ {S,T} \right\}} \right.}}}}\left( {\varphi  - r} \right)} \right)} {f_{{\gamma _{{T_1},D}}\left| T \right.}}\left( r \right)dr
\end{aligned}
\end{equation}
\setcounter{equation}{\value{mytempeqncnt}}
\hrule
\end{figure*}
where ${F_{{\gamma _{m \left| {\left\{ {S,T} \right\}} \right.}}}}\left( x  \right)$ is the CDF of $\gamma _{m}$ conditioned on $S$ and $T$ which can be written as
\setcounter{equation}{9}
\begin{equation}
\begin{aligned}
&F_{ \gamma _{{\mathop{\rm m}\nolimits} \left| {\left\{ {S,T} \right\}} \right.}}\left( x \right) = 1 - \prod\limits_{i = 1}^2 {\left( {1 - {F_{{\gamma _{{T_1},i}}|\left\{ {S,T} \right\}}}\left( x \right)} \right)} \\
&{} = 1 - \exp \left( { - \frac{{T + 1}}{{E{\Omega _1}}}x} \right)\exp \left( { - \frac{{S + {\omega _1}T + 1 + {\omega _1}}}{{{\omega _2}E{\Omega _2}}}x} \right)
\end{aligned}
\end{equation}
and ${f_{{\gamma _{T_1,D}}|T}}\left( x \right)$ is the PDF of $\gamma _{T_1,D}$ conditioned on $T$ which can be expressed as
\begin{equation}
{f_{{\gamma _{T_1,D}}|T}}\left( x \right) = \frac{{T + 1}}{{E{\Omega _0}}}\exp \left( { - \frac{{T + 1}}{{E{\Omega _0}}}x} \right).
\end{equation}
Then the CDF of $\gamma _{{T_1}}^{UB}$ can be obtained by averaging the conditioned CDF with respect to the PDFs of $S$ and $T$, i.e.,
\begin{equation}
\begin{aligned}
 {F_{\gamma _{{T_1}}^{UB}}}\left( \varphi  \right) &= \int_0^\infty  {\int_0^\infty  {{f_S}\left( s \right)} } {f_T}\left( t \right){F_{\gamma _{{T_1}}^{UB}|\left\{ {S,T} \right\}}}\left( \varphi  \right)dsdt \\
&{}  = 1 - \underbrace {\int_0^\infty  {\int_\varphi ^\infty  {{f_T}\left( t \right){f_{{\gamma _{{T_1},D}}|T}}\left(r\right)} drdt} }_{{P_1}\left( \varphi  \right)}  \\
&{}  - \int_0^\infty  {\int_0^\infty  {\int_0^\varphi  {{f_S}\left( s \right){f_T}\left( t \right){f_{{\gamma _{{T_1},D}}|T}}\left(r\right)} } }   \\
&{}  \times \left( {1 - {F_{{\gamma _{m\left| {\left\{ {S,T} \right\}} \right.}}}}\left( {\varphi  - r} \right)} \right)drdsdt,
\end{aligned}
\end{equation}
where ${{f_S}\left( s \right)}$ and ${{f_T}\left( t  \right)}$ are the PDFs of RVs $S$ and $T$, respectively. Note that $T$ is the sum of a finite number of exponential RVs with different means. Hence with the help of \cite{IEEEhowto:Khuong}, the PDF of $T$ can be written as ${f_T}\left( t \right) = \sum_{k = 1}^{{L_1}} {\frac{p_k}{E_I}\exp \big( { - \frac{t}{{{E_I}{\rho _{1,k}}}}} \big)}$, where ${\rho _{1,k}}$ is the variance of $d_{T_1,k}$ and ${p_k} = \prod_{j = 1,j \ne k}^{{L_1}} {\frac{{1}}{{{\rho _{1,k}} - {\rho _{1,j}}}}}$ for $L_1 \ge 2$ and ${p_k} = \frac{1}{{{\rho _{1,k}}}}$ for $L_1 = 1$. Substituting the expressions of ${f_T}\left( t \right)$ and ${f_{{\gamma _{T_1,D}}|T}}\left( r \right)$ into (12), we can obtain
\begin{equation}
{P_1}\left( \varphi  \right) = \exp \left( { - \frac{\varphi }{{E{\Omega _0}}}} \right)\sum\limits_j {\frac{{{p_j}{{E{\Omega _0}} \mathord{\left/
 {\vphantom {{E{\Omega _0}} {{E_I}}}} \right.
 \kern-\nulldelimiterspace} {{E_I}}}}}{{\varphi  + {{E{\Omega _0}} \mathord{\left/
 {\vphantom {{E{\Omega _0}} {{E_I}{\rho _{1,j}}}}} \right.
 \kern-\nulldelimiterspace} {{E_I}{\rho _{1,j}}}}}}} .
\end{equation}
Moreover, denoting ${\rho _{R,k}}$ as the variance of $d_{R,k}$, the PDF of $S$ can be given by ${f_S}\left( s \right) = \sum_{k = 1}^{{L_R}} {\frac{q_k}{E_I}\exp \big( { - \frac{s}{{{E_I}{\rho _{R,k}}}}} \big)}$, where ${q_k}=\prod_{j = 1,j \ne k}^{{L_R}}{\frac{1}{{{\rho _{R,k}} - {\rho _{R,j}}}}}$ for $L_R \ge 2$ and ${q_k} = \frac{1}{{{\rho _{R,k}}}}$ for $L_R=1$. By substituting the PDFs of $S$ and $T$ into (12) and using the results of Appendix, the third term in the right-hand side of (12) (denoted by $P_2\left(\varphi\right)$) can be evaluated as
%letting ${\Phi _a} = \frac{1}{{{\Omega _0}}} - \frac{1}{{{\Omega _1}}} - \frac{{{\omega _1}}}{{{\omega _2}}}\frac{1}{{{\Omega _2}}}$ and ${\Phi _b} = \frac{1}{{{\Omega _0}}} - \frac{1}{{{\Omega _1}}} - \frac{{1 + {\omega _1}}}{{{\omega _2}{\Omega _2}}}$, and according to the Appendix,
\begin{equation}
\begin{aligned}
&{P_2}\left( \varphi  \right) = \frac{{{\omega _2}}}{{E_I^2}}\exp \left( { - \frac{\varphi }{{E{\lambda _1}}}} \right)\sum\limits_j {\sum\limits_k {\frac{{{p_j}{q_k}}}{{\varphi  + {\beta _{j,k}}}}} }  \\
&{} \times \left( { {\frac{{{E^2}{\Omega _2}\lambda_2 }}{{\varphi  + {{E\lambda_2 } \mathord{\left/
 {\vphantom {{E\lambda } {{E_I}{\rho _{1,j}}}}} \right.
 \kern-\nulldelimiterspace} {{E_I}{\rho _{1,j}}}}}} - \frac{{{E^2}{\Omega _0}{\Omega _2}}}{{\varphi  + {{E{\Omega _0}} \mathord{\left/
 {\vphantom {{E{\Omega _0}} {{E_I}{\rho _{1,j}}}}} \right.
 \kern-\nulldelimiterspace} {{E_I}{\rho _{1,j}}}}}}\exp \left( { - \frac{{{\Phi _b}}}{E}\varphi } \right)} } \right.\\
  &{}+ \left. {\left( {1 + \frac{{E{\Omega _0}}}{{\varphi  + {\beta _{j,k}}}}} \right){\Theta _k}\left( \varphi  \right) + \left( {\frac{1}{{{\omega _2}}} + \frac{{E{\Omega _0}{\Omega _2}{\Phi _a}}}{{\varphi  + {\beta _{j,k}}}}} \right){\Xi _j}\left( \varphi  \right)} \right),
\end{aligned}
\end{equation}
where ${\Phi _a} = \frac{1}{{{\Omega _0}}} - \frac{1}{{{\Omega _1}}} - \frac{{{\omega _1}}}{{{\omega _2}}}\frac{1}{{{\Omega _2}}}\ne0$, ${\Phi _b} = \frac{1}{{{\Omega _0}}} - \frac{1}{{{\Omega _1}}} - \frac{{1 + {\omega _1}}}{{{\omega _2}{\Omega _2}}}$, ${\beta _{j,k}} = \frac{{E{\Omega _0}}}{{{E_I}}}( \frac{1}{{{\rho _{1,j}}}} +\frac{{{\omega _2}{\Phi _a}{\Omega _2}}}{{{\rho _{R,k}}}} )$, $\lambda_1  = {( {\frac{1}{{{\Omega _1}}} + \frac{{{\omega _1+1}}}{{{\omega _2}}}\frac{1}{{{\Omega _2}}}})^{ - 1}}$ and $\lambda_2  = {( {\frac{1}{{{\Omega _1}}} + \frac{{{\omega _1}}}{{{\omega _2}}}\frac{1}{{{\Omega _2}}}})^{ - 1}}$. ${\Theta _k}\left( \varphi  \right)$ and ${\Xi _j}\left( \varphi  \right)$ can be {expressed\footnotemark[4]} as in (15) and (16) at the top of the next page, where ${\vartheta _j} = \frac{\varphi }{{{\Phi _a}}}( {\frac{1}{{{\Omega _1}}} + \frac{{{\omega _1}}}{{{\omega _2}}}\frac{1}{{{\Omega _2}}}} ) + \frac{E}{{{\Phi _a}{E_I}{\rho _{1,j}}}}$. $\textrm{Ei}(\cdot)$ and  $\phi \left( \cdot \right)$ are the exponential integral [13, 3.351.6] and lower incomplete gamma function [13, 8.350.1], respectively. Besides, for the case of ${\Phi _a} = 0$, $ P_2\left(\varphi\right)$ can be written as [Appendix]
{\footnotetext[4]{Note that the series expression of ${\Theta _k}\left( \varphi  \right)$ (for $\Phi _b>0$ in (15)) is also valid for the case of $\Phi _b<0$. However, we present another closed-form expression without infinite series to facilitate the computation of ${\Theta _k}\left( \varphi  \right)$ when $\Phi _b<0$. Similarly, it can be seen the series expression of ${\Xi _j}\left( \varphi  \right)$ (for $\Phi _b<0<\Phi _a$ in (16)) is always valid except for the case ($\Phi _a=0$ or $\Phi _b=0$).}}
\begin{figure*}
\normalsize
\setcounter{mytempeqncnt}{\value{equation}}
% Set the equation number to one less than the one
% desired for the first equation here.
% The value here will have to changed if equations
% are added or removed prior to the place these
% equations are referenced in the main text.\int\limits_{}^{} {}
%\quad\;
\setcounter{equation}{14}
\begin{equation}
{\Theta _k}\left( \varphi  \right) = \left\{ \begin{aligned}
& \sum\limits_{l = 0}^\infty  {\frac{{E{\Omega _2}}}{{{{\left( {\varphi  + {{{\omega _2}E{\Omega _2}} \mathord{\left/
 {\vphantom {{{\omega _2}E{\Omega _2}} {{E_I}{\rho _{R,k}}}}} \right.
 \kern-\nulldelimiterspace} {{E_I}{\rho _{R,k}}}}} \right)}^{l + 1}}}}} {\left( {\frac{{{\Phi _b}}}{E}} \right)^{ - \left( {l + 1} \right)}}\phi \left( {l + 1,\frac{{\varphi {\Phi _b}}}{E}} \right),{\Phi _b} > 0  \\
&{} E{\Omega _2}\exp \left( { - \frac{{{\Phi _b}}}{E}\left[ {\varphi  + \frac{{{\omega _2}E{\Omega _2}}}{{{E_I}{\rho _{R,k}}}}} \right]} \right)\left( {{\rm{Ei}}\left( {\frac{{{\Phi _b}}}{E}\left[ {\varphi  + \frac{{{\omega _2}E{\Omega _2}}}{{{E_I}{\rho _{R,k}}}}} \right]} \right) - {\rm{Ei}}\left( {\frac{{{\omega _2}{\Phi _b}{\Omega _2}}}{{{E_I}{\rho _{R,k}}}}} \right)} \right),{\Phi _b} < 0 \\
&{} E{\Omega _2}\ln \left( {\frac{{{E_I}{\rho _{R,k}}}}{{{\omega _2}E{\Omega _2}}}\varphi  + 1} \right),{\Phi _b} = 0\\
\end{aligned} \right.
\end{equation}
\setcounter{equation}{\value{mytempeqncnt}}
\hrule
\end{figure*}
\begin{figure*}
\normalsize
\setcounter{mytempeqncnt}{\value{equation}}
% Set the equation number to one less than the one
% desired for the first equation here.
% The value here will have to changed if equations
% are added or removed prior to the place these
% equations are referenced in the main text.\int\limits_{}^{} {}
%\quad\;\textrm{Ei}\left( {-\frac{{{\Phi _b}}}{E}\frac{{\varphi  + \frac{{E{\Omega _0}}}{{{E_I}{\rho _{1,j}}}}}}{{{\Omega _0}{\Phi _a}}}} \right)}
\setcounter{equation}{15}
\begin{equation}
{\Xi _j}\left( \varphi  \right) = \left\{ \begin{aligned}
 & \frac{E}{{{\Phi _a}}}\exp \left( {\frac{{{\Phi _b}{\vartheta _j}}}{E}} \right)\left( {\textrm{Ei}\left( { - \frac{{{\Phi _b}\left( {\varphi  + {\vartheta _j}} \right)}}{E}} \right) - \textrm{Ei}\left( { - \frac{{{\Phi _b}{\vartheta _j}}}{E}} \right)} \right),{\Phi _a} > {\Phi _b} > 0 \\
 &{} \exp \left( { - \frac{{{\Phi _b}\varphi }}{E}} \right)\sum\limits_{l = 0}^\infty  {\frac{{E\Phi _a^l\Omega _0^{l + 1}}}{{{{\left( {\varphi  + {{E{\Omega _0}} \mathord{\left/
 {\vphantom {{E{\Omega _0}} {{E_I}{\rho _{1,j}}}}} \right.
 \kern-\nulldelimiterspace} {{E_I}{\rho _{1,j}}}}} \right)}^{l + 1}}}}} {\left( { - \frac{{{\Phi _b}}}{E}} \right)^{ - \left( {l + 1} \right)}}\phi \left( {l + 1,-\frac{{\varphi {\Phi _b}}}{E}} \right) ,{\Phi _b} < 0 < {\Phi _a} \\
  &{}  \frac{E}{{{\Phi _a}}}\exp \left( {-\frac{{{\Phi _b}}}{E}\left[ {\varphi  - \frac{{\varphi  + \frac{{E{\Omega _0}}}{{{E_I}{\rho _{1,j}}}}}}{{{\Omega _0}{\Phi _a}}}} \right]} \right)\left( {\textrm{Ei}\left( {-\frac{{{\Phi _b}}}{E}\frac{{\varphi  + \frac{{E{\Omega _0}}}{{{E_I}{\rho _{1,j}}}}}}{{{\Omega _0}{\Phi _a}}}} \right) - {\rm{Ei}}\left( {\frac{{{\Phi _b}}}{E}\left[ {\varphi  - \frac{{\varphi  + \frac{{E{\Omega _0}}}{{{E_I}{\rho _{1,j}}}}}}{{{\Omega _0}{\Phi _a}}}} \right]} \right)} \right),{\Phi _b} < {\Phi _a} < 0 \\
 &{}\frac{E}{{{\Phi _a}}}\ln \left( {\frac{{1 + {{\varphi {E_I}{\rho _{1,j}}} \mathord{\left/
 {\vphantom {{\varphi {E_I}{\rho _{1,j}}} {E{\Omega _0}}}} \right.
 \kern-\nulldelimiterspace} {E{\Omega _0}}}}}{{1 + {{\varphi {E_I}{\rho _{1,j}}} \mathord{\left/
 {\vphantom {{\varphi {E_I}{\rho _{1,j}}} {\lambda E}}} \right.
 \kern-\nulldelimiterspace} {\lambda E}}}}} \right),{\Phi _b} = 0 \\
 \end{aligned} \right.
\end{equation}
\setcounter{equation}{\value{mytempeqncnt}}
\hrule
\end{figure*}
\setcounter{equation}{16}
\begin{equation}
\begin{aligned}
&{P_2}\left(\varphi\right)   = \frac{{{\omega _2}}}{{E_I^2}}\exp \left( { - \frac{\varphi }{{E{\lambda _1}}}} \right)\sum\limits_j {\sum\limits_k {{p_j}} } {q_k}\\
&{}  \times\! \frac{1}{{\varphi  \!+\! {{E{\Omega _0}} \mathord{\left/
 {\vphantom {{E{\Omega _0}} {{E_I}{\rho _{1,j}}}}} \right.
 \kern-\nulldelimiterspace} {{E_I}{\rho _{1,j}}}}}}\left( {1 \!+\! \frac{{E{\Omega _0}}}{{\varphi  + {{E{\Omega _0}} \mathord{\left/
 {\vphantom {{E{\Omega _0}} {{E_I}{\rho _{1,j}}}}} \right.
 \kern-\nulldelimiterspace} {{E_I}{\rho _{1,j}}}}}}} \right)\!{\Theta _k}\left( \varphi  \right).
\end{aligned}
\end{equation}
Then we will test the convergence of the infinite series involved in the expressions of ${\Theta _k}\left( \varphi  \right)$ and ${\Xi _j}\left( \varphi  \right)$. Defining
\begin{equation}
{\Delta _l}\left( {{\eta _1},{\eta _2},{\eta _3},{\eta _4}} \right) = \frac{{\eta _1^l\eta _4^{ - \left( {l + 1} \right)}}}{{{{\left( {{\eta _2}\varphi  + {\eta _3}} \right)}^{l + 1}}}}\phi \left( {l + 1,\varphi {\eta _4}} \right),
\end{equation}
where $0\!\! <\!\!{\eta _1}\!\! \le\!\! {\eta _2}$ and $\eta _3>0$. Then it can be shown that ${\Theta _k}\left( \varphi  \right) = {\Delta _l}\big( {\frac{1}{{E{\Omega _2}}},\frac{1}{{E{\Omega _2}}},\frac{\omega_2}{{{E_I}{\rho _{R,k}}}},\frac{{{\Phi _b}}}{E}} \big)$ $\left({\Phi _b} > 0 \right)$ and ${\Xi _j}\left( \varphi  \right) = \exp \left( { - \frac{{{\Phi _b}}}{E}\varphi } \right){\Delta _l}\big( {\frac{{{\Phi _a}}}{E},\frac{1}{{E{\Omega _0}}},\frac{1}{{{E_I}{\rho _{1,j}}}}, - \frac{{{\Phi _b}}}{E}} \big)$ $\left({\Phi _b}  \right.$$\left.<0< {\Phi _a}\right)$, thus it is sufficient to prove that the infinite series $\sum_{l = 1}^\infty  {{\Delta _l}\left( {{\eta _1},{\eta _2},{\eta _3},{\eta _4}} \right)}$ is convergent. Using [13, 3.381.1], it can be shown that
\begin{equation}
\begin{aligned}
& \begin{array}{l}
 \mathop {\lim }\limits_{l \to \infty } \sqrt[l]{{{\Delta _l}\left( {{\eta _1},{\eta _2},{\eta _3},{\eta _4}} \right)}}\\
  = \mathop {\lim }\limits_{l \to \infty } \sqrt[l]{{\frac{{\eta _1^l}}{{{{\left( {{\eta _2}\varphi  + {\eta _3}} \right)}^{l + 1}}}}\int_0^\varphi  {{r^l}\exp \left( { - {\eta _4}r} \right)} dr}} \\
   \le \mathop {\lim }\limits_{l \to \infty } \sqrt[l]{{\frac{{\eta _1^l{\varphi ^l}}}{{{{\left( {{\eta _2}\varphi  + {\eta _3}} \right)}^{l + 1}}}}\int_0^\varphi  {\exp \left( { - {\eta _4}r} \right)} dr}}\\
   \end{array}\\
   &{}= \frac{{{\eta _1}\varphi }}{{{\eta _2}\varphi  + {\eta _3}}} < 1
   \end{aligned}
\end{equation}
By the root test \cite{IEEEhowto:Rudin}, it can be seen that the infinite series in (15) and (16) are always convergent when $\varphi<\infty$.

Finally, the lower bound of the outage probability for the AF-based TDBC protocol in the presence of CCI can be derived by substituting (13) and (14) (or (17)) in to (12), i.e.,
\begin{equation}
P_{{T_1}}^{OUT-LB}\left( {{R_t}} \right) = {F_{\gamma _{{T_1}}^{UB}}}\left( \varphi  \right) = 1 - P_1\left( \varphi  \right) - P_2\left( \varphi  \right).
\end{equation}
\subsection{Asymptotic Analysis}
To offer an intuitive observation into the effect of CCI on the outage performance, we develop asymptotic analysis on the outage probability based on the analysis of subsection A. According to \cite{IEEEhowto:Wang}\cite{IEEEhowto:Suraweera}, the asymptotic expression can be derived by performing McLaurin series expansion to ${F_{\gamma _{{T_1}}^{UB}}}\left( \varphi  \right)$ and taking only the first two order terms. Wherein the McLaurin series expansion of ${\Theta _k}\left( \varphi  \right)$ can be given by
\begin{equation}
{\Theta _k}\left( \varphi  \right) = {\Theta _k}\left( 0 \right) + \Theta _k^{\left( 1 \right)}\left( 0 \right)\varphi  + \frac{{\Theta _k^{\left( 2 \right)}\left( 0 \right)}}{2}{\varphi ^2} + {\cal O}\left( {{\varphi ^2}} \right), \\
\end{equation}
where ${\cal O}\left( \delta  \right)$ indicates the higher order term of $\delta$ and $\Theta _k^{\left( n \right)}\left( 0 \right)$ ($n\!=\!1,2$) can be determined {by\footnotemark[5]} [13, 0.410]:
{\footnotetext[5]{Note the $\Theta _k^{\left( n \right)}\left( \varphi \right)$ here is obtained by taking the derivative of its integral expression which is given by (26).}}
\begin{equation}
\begin{aligned}
& \Theta _k^{\left( 1 \right)}\left( 0 \right) = {\left[ {{{\left. {g\left( {\varphi ,r} \right)} \right|}_{r = \varphi }}} \right]_{\varphi  = 0}}\\
&{} \Theta _k^{\left( 2 \right)}\left( 0 \right) = {\left[ {{{\left. {\frac{{dg\left( {\varphi ,r} \right)}}{{d\varphi }}} \right|}_{r = \varphi }} + \frac{d}{{d\varphi }}\left( {{{\left. {g\left( {\varphi ,r} \right)} \right|}_{r = \varphi }}} \right)} \right]_{\varphi  = 0}},
\end{aligned}
\end{equation}
where $g\left( {\varphi ,r} \right) = {\big( {\frac{{\varphi  - r}}{{E{\Omega _2}}} + \frac{{{\omega _2}}}{{{E_I}{\rho _{R,k}}}}} \big)^{ - 1}}\exp \left( { - \frac{{{\Phi _b}}}{E}r} \right)$. Moreover, the McLaurin series expansion of ${\Xi _j}\left( \varphi  \right)$ can be calculated using the similar method as in the above. Finally, the asymptotic expression of ${F_{\gamma _{{T_1}}^{UB}}}\left( \varphi  \right)$ can be written {as\footnotemark[6]}
{\footnotetext[6]{The asymptotic expression here is obtained based on the expression of ${F_{\gamma _{{T_1}}^{UB}}}\left( \varphi  \right)$ in the case of $\Phi _a\ne0$, however, for the case of $\Phi _a=0$, it can be verified that this expression is also valid.}}
\begin{equation}
\begin{aligned}
{F_{\gamma _{{T_1}}^{UB}}}\left( \varphi  \right) & \approx  {\left( {\frac{{{E_I}}}{E}} \right)^2}\frac{{{\varphi ^2}}}{{2{\Omega _0}}}\left[ {\sum\limits_j {\sum\limits_k {\frac{{{p_j}{q_k}\rho _{R,k}^2}}{{{\omega _2}{\Omega _2}}}} } \left( {\frac{{{\rho _{1,j}}}}{{{E_I}}} + \rho _{1,j}^2} \right)} \right. \\
 &{} \left. {\sum\limits_j {{p_j}} \left( {\frac{{{\rho _{1,j}}}}{{E_I^2{\lambda _1}}} + \left( {\frac{1}{{{\lambda _1}}} + \frac{1}{{{\lambda _2}}}} \right)\frac{{\rho _{1,j}^2}}{{{E_I}}} + \frac{{2\rho _{1,j}^3}}{{{\lambda _2}}}} \right)} \right]. \\
 \end{aligned}
\end{equation}
Through the asymptotic expression, it can be seen that when the ratio of useful power to interference power is constant, the AF-based TDBC protocol dose not achieve any diversity.
\section{Simulation Results}
In this section, we provide the simulation results to verify our theoretical analyses on the outage probability. It is assumed that $T_1$, $T_2$ and $R$ are located in a straight line and $R$ is between $T_1$ and $T_2$. The distance between two terminals is normalized to 1 and the path loss exponent is set to 4 \cite{IEEEhowto:Rappaport}, thus the variances of $h_0$, $h_1$ and $h_2$ can be computed as ${\Omega _0}=1$, ${\Omega _1}=D_1^{-4}$ and ${\Omega _2}=\left(1-D_1\right)^{-4}$, respectively, where $D_1$ indicates the normalized distance between $T_1$ and $R$. The normalized distances between node $N$ and the interferers that interfere $N$ are assumed to be evenly distributed on the interval $\left(\alpha_1 ,\alpha_2 \right)=(1,1.5)$, where $N \in \left\{ T_1,T_2,R\right\}$. Hence, ${\rho _{R,k}}$ and ${\rho _{i,k}}$ can be determined by ${\rho _{R,k}} = \left( {\alpha_1  + \left( {k - 1} \right)\left( {\alpha_2  - \alpha_1 } \right)}/\left(L_R-1\right) \right)^{-4}$ and ${\rho _{i,k}} = \left( {\alpha_1  + \left( {k - 1} \right)\left( {\alpha_2  - \alpha_1 } \right)}/\left(L_i-1\right) \right)^{-4}$.
\begin{figure}[t]
\centering
\includegraphics[width=7.43cm]{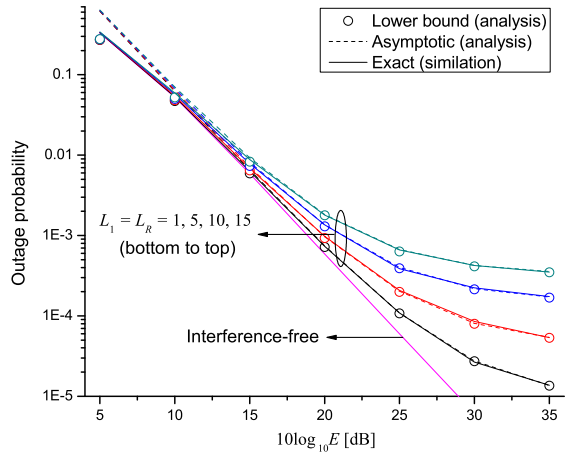}
\caption{Outage performance at $T_1$ with fixed $E/E_I$, where $D_1$ = 0.5, ${\omega _1}=0.5$ and $R_t=1$ bit/s/Hz.} \label{fig:graph}
\end{figure}
\begin{figure}[t]
\centering
\includegraphics[width=7.43cm]{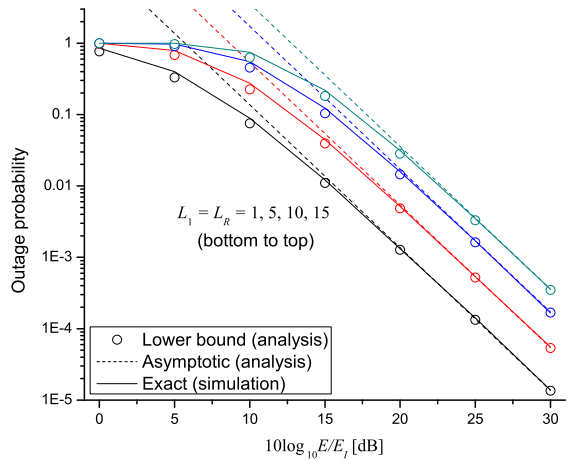}
\caption{Outage performance at $T_1$ versus $E/E_I$, where $E_I=5$dB, $D_1$ = 0.5, ${\omega _1}=0.5$ and $R_t=1$ bit/s/Hz.} \label{fig:graph}
\end{figure}

In Fig. 2, the outage performance at $T_1$ is presented as a function of the transmit power $E$, where $E/E_I$ is fixed at 30dB. The outage performance of the scenario without interference is also presented as a benchmark. It is seen that the outage performance degrades as the numbers of interferers increase. Furthermore, the slope of outage probability curves are steep in the low SNR region ($E<15$dB). This is because the power of AWGN dominates the interference power. While a performance floor can be observed in the high SNR region due to the dominant role of interference power. This phenomenon also indicates that the achievable diversity order of the TDBC protocol in the interference-limited scenario is zero. Fig. 3 studies the outage performance at $T_1$ against $E/E_I$. From Fig. 3, it can be seen that the outage probability increases as the numbers of interferers as well as $E_I/E$ increase as expected. Finally, a fine agreement between the analytic results and simulations can be observed from the figures.

Fig. 4 investigates the effect of relay placement on the outage performance, where numbers of interferers that affect $T_1$ and $R$ may different. Without loss of generality, we set $L_1=1$ and let $L_R$ increase from 1, 5, 10 to 15. Then we examine the outage performance at $T_1$ as a function of $D_1$. It is seen from the figure that the optimal relay placement moves toward $T_2$ as the number of interferers (and total interference power) that affects the relay increases. This is because the AF operation adopted by the relay. From equation (4), we can see that, for given $\left(\alpha_1,\alpha_2\right)$, when $L_R$ increases, to decrease the amplified interference (i.e., the term $\left( {{\cal A}_1^2 + {\cal A}_2^2} \right)E{{\left| {{h_1}} \right|}^2}\big( {{E_I}\sum_{k = 1}^{{L_R}} {{{\left| {{d_{R,k}}} \right|}^2} + 1} }\big)$), the relay should move toward $T_2$ to decrease $\left|h_1\right|^2$.

\section{Conclusions}
In this paper, we study the effect of CCI on the AF-based TDBC protocol. Lower bound of the outage probability is derived and is shown to provide a good match with the simulation results. Meanwhile, a simpler asymptotic expression of outage probability is also provided. We show through both analytic and simulation results that the achievable diversity of the TDBC protocol in the interference limited scenario is zero. Moreover, we investigate the effect of relay placement on the outage probability and show that when only consider the outage performance at one terminal (e.g. $T_1$), as the number of interferers that interferes the relay increases, the optimal relay placement needs to move toward $T_2$ in order to obtain the optimal outage probability at $T_1$.

\appendix
\section{}
Substituting the PDFs of $S$ and $T$ into (12) and interchanging the integration order, we can obtain
\begin{equation}
\begin{aligned}
& {P_2}\left( \varphi  \right) = \frac{1}{{E_I^2}}\exp \left( { - \left( {\frac{1}{{{\Omega _1}}} + \frac{{{\omega _1} + 1}}{{{\omega _2}{\Omega _2}}}} \right)\frac{\varphi }{E}} \right)\sum\limits_j {\sum\limits_k {\frac{{{p_j}{q_k}}}{{E{\Omega _0}}}} }    \\
&{} \int_0^\varphi  {\exp \left( { - \frac{{{\Phi _b}}}{E}r} \right)\int_0^\infty  {\exp \left( { - \left[ {\frac{{\varphi  - r}}{{{\omega _2}E{\Omega _2}}} + \frac{1}{{{E_I}{\rho _{R,k}}}}} \right]s} \right)} ds}    \\
&{} \int_0^\infty  {\exp \left( { - \left[ {\frac{{{\Phi _a}r}}{E} + \left( {\frac{1}{{{\Omega _1}}} + \frac{{{\omega _1}}}{{{\omega _2}}}\frac{1}{{{\Omega _2}}}} \right)\frac{\varphi }{E} + \frac{1}{{{E_I}{\rho _{1,j}}}}} \right]t} \right)}  \\
&{} \times \left( {t + 1} \right)dtdr.
\end{aligned}
\end{equation}
\emph{Case1} $\left( {\Phi _a} \ne 0 \right)$: In this case, solving the integrals with respect to $s$ and $t$, we can yield
\begin{equation}
\begin{array}{l}
{P_2}\left( \varphi  \right) = \frac{{{\omega _2}}}{{E_I^2}}\exp \left( { - \left( {\frac{1}{{{\Omega _1}}} + \frac{{{\omega _1} + 1}}{{{\omega _2}{\Omega _2}}}} \right)\frac{\varphi }{E}} \right) \\
\times \sum\limits_j  \sum\limits_k  \frac{{{p_j}{q_k}}}{{\varphi  + \frac{{E{\Omega _0}}}{{{E_I}}}\left( {\frac{1}{{{\rho _{1,j}}}} + \frac{{{\omega _2}{\Phi _a}{\Omega _2}}}{{{\rho _{R,k}}}}} \right)}}
\\
\times \left\{\left[ { { {\frac{{E{\Omega _2}}}{{\frac{\varphi }{E}\left( {\frac{1}{{{\Omega _1}}} + \frac{{{\omega _1}}}{{{\omega _2}}}\frac{1}{{{\Omega _2}}}} \right) + \frac{1}{{{E_I}{\rho _{1,j}}}}}} - \frac{{E{\Omega _2}}}{{\frac{\varphi }{{E{\Omega _0}}} + \frac{1}{{{E_I}{\rho _{1,j}}}}}}\exp \left( { - \frac{{{\Phi _b}}}{E}\varphi } \right)} }} \right]\right. \\
+ \left( {1 + \frac{{E{\Omega _0}}}{{\varphi  + \frac{{E{\Omega _0}}}{{{E_I}}}\big( {\frac{1}{{{\rho _{1,j}}}} + \frac{{{\omega _2}{\Phi _a}{\Omega _2}}}{{{\rho _{R,k}}}}} \big)}}} \right){\Theta _k}\left( \varphi  \right) \\
\left. { + \left( {\frac{1}{{{\omega _2}}} + \frac{{E{\Omega _0}{\Omega _2}{\Phi _a}}}{{\varphi  + \frac{{E{\Omega _0}}}{{{E_I}}}\big( {\frac{1}{{{\rho _{1,j}}}} + \frac{{{\omega _2}{\Phi _a}{\Omega _2}}}{{{\rho _{R,k}}}}} \big)}}} \right){\Xi _j}\left( \varphi  \right)} \right\},
 \end{array}
\end{equation}
where ${\Theta _k}\left( \varphi  \right)$ and ${\Xi _j}\left( \varphi  \right)$ are expressed as
\begin{equation}
\begin{array}{l}
\!\!\!\!\!\!{\Theta _k}\left( \varphi  \right) \!=\! \int_0^\varphi\!  {\frac{1}{{\frac{{\varphi  - r}}{{E{\Omega _2}}} + \frac{{{\omega _2}}}{{{E_I}{\rho _{R,k}}}}}}\exp \left( { - \frac{{{\Phi _b}}}{E}r} \right)} dr\\
\!\!\!\!\!\!{\Xi _j}\left( \varphi  \right) \!=\! \int_0^\varphi \!  \frac{1}{{\frac{{{\Phi _a}}}{E}r + \left( {\frac{1}{{{\Omega _1}}} + \frac{{{\omega _1}}}{{{\omega _2}}}\frac{1}{{{\Omega _2}}}} \right)\frac{\varphi }{E} + \frac{1}{{{E_I}{\rho _{1,j}}}}}}\exp \left( { - \frac{{{\Phi _b}}}{E}r} \right) dr
\end{array}
\end{equation}
When ${\Phi _b} > 0$, to solve the integral ${\Theta _k}\left( \varphi  \right)$, we apply Taylor series expansion ${\big( {\frac{{\varphi  - r}}{{E{\Omega _2}}} + \frac{{{\omega _2}}}{{{E_I}{\rho _{R,k}}}}} \big)^{ - 1}} = E{\Omega _2}\sum_{l = 0}^\infty  {{{{r^l}} \mathord{\left/
 {\vphantom {{{r^l}} {{{\left( {\varphi  + {{{\omega _2}E{\Omega _2}} \mathord{\left/
 {\vphantom {{{\omega _2}E{\Omega _2}} {{E_I}{\rho _{R,k}}}}} \right.
 \kern-\nulldelimiterspace} {{E_I}{\rho _{R,k}}}}} \right)}^{l + 1}}}}} \right.
 \kern-\nulldelimiterspace} {{{\left( {\varphi  + {{{\omega _2}E{\Omega _2}} \mathord{\left/
 {\vphantom {{{\omega _2}E{\Omega _2}} {{E_I}{\rho _{R,k}}}}} \right.
 \kern-\nulldelimiterspace} {{E_I}{\rho _{R,k}}}}} \right)}^{l + 1}}}}} $. Then based on [13, 3.381.1], the integral can be solved into
\begin{equation}
{\Theta _k}\left( \varphi  \right){\rm{ }} \!=\! \sum\limits_{l = 0}^\infty  {\frac{{E{\Omega _2}}}{{{{\left( {\varphi  + \frac{{{\omega _2}E{\Omega _2}}}{{{E_I}{\rho _{R,k}}}}} \right)}^{l + 1}}}}} {\left( {\frac{{{\Phi _b}}}{E}} \right)^{ - \left( {l + 1} \right)}}\!\phi \left( {l + 1,\frac{{\varphi {\Phi _b}}}{E}} \right).
\end{equation}
When ${\Phi _b} < 0$, ${\Theta _k}\left( \varphi  \right)$ can be solved by replacing $\left(\varphi  - r\right)$ with $t$, and then using the integral result reported in [13, 3.352.1],
\begin{equation}
\begin{aligned}
{\Theta _k}\left( \varphi  \right) &= \exp \left( { - \frac{{{\Phi _b}}}{E}\varphi } \right)\int_0^\varphi  {\frac{1}{{\frac{t}{{E{\Omega _2}}} + \frac{{{\omega _2}}}{{{E_I}{\rho _{R,k}}}}}}\exp \left( {\frac{{{\Phi _b}}}{E}t} \right)} dt \\
&{} = E{\Omega _2}\exp \left( { - \frac{{{\Phi _b}}}{E}\left[ {\varphi  + \frac{{{\omega _2}E{\Omega _2}}}{{{E_I}{\rho _{R,k}}}}} \right]} \right) \\
&{} \times \left( {{\rm{Ei}}\left( {\frac{{{\Phi _b}}}{E}\left[ {\varphi  + \frac{{{\omega _2}E{\Omega _2}}}{{{E_I}{\rho _{R,k}}}}} \right]} \right) - {\rm{Ei}}\left( {\frac{{{\omega _2}{\Phi _b}{\Omega _2}}}{{{E_I}{\rho _{R,k}}}}} \right)} \right).
\end{aligned}
\end{equation}
On the other hand, using [13, 3.352.1] on ${\Xi _j}\left( \varphi  \right)$ when ${\Phi _a} > {\Phi _b} > 0$, we can yield ${\Xi _j}\left( \varphi  \right) = \frac{E}{{{\Phi _a}}}\exp \big( {\frac{{{\Phi _b}{\vartheta _j}}}{E}} \big)\big( {\textrm{Ei}\big( { - \frac{{{\Phi _b}\left( {\varphi  + {\vartheta _j}} \right)}}{E}} \big) - \textrm{Ei}\big( { - \frac{{{\Phi _b}{\vartheta _j}}}{E}} \big)} \big)$. To solve the integral ${\Xi _j}\left( \varphi  \right)$ when ${\Phi _b} < 0 < {\Phi _a}$, similar approach as in the case of ${\Phi _b} > 0$ for  ${\Theta _k}\left( \varphi  \right)$ can be used, then we obtain
\begin{equation}
\begin{aligned}
{\Xi _j}\left( \varphi  \right)&\mathop  = \limits^{\varphi  - r = t} \exp \left( \! { - \frac{{{\Phi _b}\varphi }}{E}}\! \right)\!\! \int_0^\varphi \!\!  {\frac{1}{{\frac{\varphi }{{E{\Omega _0}}}\! +\! \frac{1}{{{E_I}{\rho _{1,j}}}}\! -\! \frac{{{\Phi _a}}}{E}t}}\exp \!\! \left( \!{\frac{{{\Phi _b}}}{E}t}\! \right)} dt\\
&{}= \exp \left( { - \frac{{{\Phi _b}\varphi }}{E}} \right)\sum\limits_{l = 0}^\infty  {\frac{{E\Phi _a^l\Omega _0^{l + 1}}}{{{{\left( {\varphi  + {{E{\Omega _0}} \mathord{\left/
 {\vphantom {{E{\Omega _0}} {{E_I}{\rho _{1,j}}}}} \right.
 \kern-\nulldelimiterspace} {{E_I}{\rho _{1,j}}}}} \right)}^{l + 1}}}}} \\
&{} \times {\left( { - \frac{{{\Phi _b}}}{E}} \right)^{ - \left( {l + 1} \right)}}\phi \left( {l + 1,-\frac{{\varphi {\Phi _b}}}{E}} \right).
\end{aligned}
\end{equation}
Using the similar method as in the case of ${\Phi _b} < 0$ for ${\Theta _k}\left( \varphi  \right)$, it can be shown that, when $ {\Phi _b}< {\Phi _a} < 0$,
\begin{equation}
\begin{aligned}
& {\Xi _j}\left( \varphi  \right) =  - \frac{E}{{{\Phi _a}}}\exp \left( { - \frac{{{\Phi _b}}}{E}\left[ {\varphi  - \frac{{\varphi  + \frac{{E{\Omega _0}}}{{{E_I}{\rho _{1,j}}}}}}{{{\Omega _0}{\Phi _a}}}} \right]} \right) \\
&{} \times \! \left( {\textrm{Ei}\left( {\frac{{{\Phi _b}}}{E}\left[ {\varphi  - \frac{{\varphi  + \frac{{E{\Omega _0}}}{{{E_I}{\rho _{1,j}}}}}}{{{\Omega _0}{\Phi _a}}}} \right]} \right) \!-\! \textrm{Ei}\left( {-\frac{{{\Phi _b}}}{E}\frac{{\varphi  + \frac{{E{\Omega _0}}}{{{E_I}{\rho _{1,j}}}}}}{{{\Omega _0}{\Phi _a}}}} \right)} \right).
\end{aligned}
\end{equation}
Moreover, it is very easy to verify that ${\Theta _k}\left( \varphi  \right)$ and ${\Xi _j}\left( \varphi  \right)$ can be expressed as in (15) and (16) when ${\Phi _b} = 0$, thus the derivations details for this case are omitted.

\emph{Case2} $\left( {\Phi _a} = 0 \right)$: In this case, it is easy to verify by using the equation $\Phi_a=0$ that the sum of the first term and third term in bracket $\{\cdot\}$ of (25) equals to zero, thus $P_2\left(\varphi\right)$ can be simplified to (17).

\begin{figure}[t]
\centering
\includegraphics[width=7.43cm]{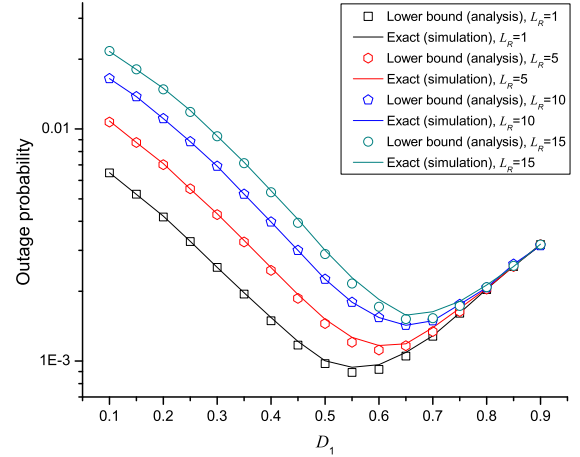}
\caption{Outage performance at $T_1$ versus relay placement, $E\!=\!30$dB, $E_I\!=\!10$dB, ${\omega _1}\!=\!0.5$, $R_t\!=\!1$ bit/s/Hz and $L_1\!=\!1$.} \label{fig:graph}
\end{figure}

\section*{Acknowledgment}
This work is supported by the Jiangsu Province Natural Science Foundation (BK2011002), Major Special Project of China (2010ZX03003-003-01), National Natural Science Foundation of China (No. 60972050) and Jiangsu Province Natural Science Foundation for Young Scholar (BK2012055).

\end{document}